\input epsf
\documentclass[nohyper,notoc]{article} 
\usepackage{epsfig}
\usepackage{bm}
\usepackage{dcolumn}
\usepackage{latexsym}

\def\bit{\begin{itemize}}
\def\eit{\end{itemize}}
\def\ben{\begin{enumerate}}
\def\een{\end{enumerate}}
\def\beq{\begin{equation}}
\def\eeq{\end{equation}}
\def\bea{\begin{eqnarray}}
\def\eea{\end{eqnarray}}
\def\bq{\begin{quote}}
\def\eq{\end{quote}}
\def \lsim{\mathrel{\vcenter
     {\hbox{$<$}\nointerlineskip\hbox{$\sim$}}}}
\def \gsim{\mathrel{\vcenter
     {\hbox{$>$}\nointerlineskip\hbox{$\sim$}}}}
\def\gappeq{\mathrel{\rlap {\raise.5ex\hbox{$>$}}
{\lower.5ex\hbox{$\sim$}}}}
\def\lappeq{\mathrel{\rlap{\raise.5ex\hbox{$<$}}
{\lower.5ex\hbox{$\sim$}}}}

\def\meg{\mu \to e \gamma}
\def\teg{\tau \to e \gamma}
\def\tmg{\tau \to \mu \gamma}
\def\m3e{\mu \to e \bar{e} e}

\def\Ztl{Z \to \tau^\pm  \ell^\mp }
\def\Ztm{Z \to \tau^\pm  \mu^\mp }
\def\Zme{Z \to \mu^\pm  e^\mp }
\def\Zte{Z \to \tau^\pm  e^\mp }
\def\Ztt{Z \to \tau^\pm  \tau^\mp }
\def\Zmm{Z \to \mu^\pm  \mu^\mp }

\def\tennb{\tau^\pm \to e^\pm \nu \bar{\nu}}

\def\tmnnb{\tau^\pm \to \mu^\pm \nu \bar{\nu}}
\def\tlg{\tau \to \ell \gamma}

\def\meee{\mu \to e \bar{e} e}
\def\teee{\tau \to e \bar{e} e}
\def\tlll{\tau \to \ell \bar{\ell} \ell}
\def\tmmm{\tau \to \mu \bar{\mu} \mu}

\def\a{\alpha}
\def\b{\beta}
\def\g{\gamma}

\def\m{\mu}

\newcommand{\met}{\mbox{\ensuremath{\slash\kern-.7emE_{T}}}}

\begin{document}
\vspace*{-1in}
\renewcommand{\thefootnote}{\fnsymbol{footnote}}
\vskip 5pt
\begin{center}
{\Large {\bf 
LHC sensitivity to  lepton flavour violating Z boson decays
}}
\vskip 25pt
{\bf   Sacha Davidson $^{1,}$\footnote{E-mail address:
s.davidson@ipnl.in2p3.fr}, 
Sylvain Lacroix $^{1,}$\footnote{E-mail address:
s.lacroix@ipnl.in2p3.fr}
 and Patrice Verdier $^{1,}$}\footnote{E-mail address:
verdier@ipnl.in2p3.fr} 
 
\vskip 10pt  
$^1${\it IPNL, Universit\'e de Lyon, Universit\'e Lyon 1, CNRS/IN2P3, 4 rue E. Fermi 69622 Villeurbanne cedex, France }\\
\vskip 20pt
{\bf Abstract}
\end{center}

\begin{quotation}
  {\noindent\small

We estimate  that the LHC could  set bounds 
  $BR(Z\to \mu^\pm e^\mp) < 4.1\times 10^{-7}$ and
  $BR(Z \to \tau^\pm \mu^\mp)< 3.5\times 10^{-6}$
(at 95\% C.L.)
with 20 fb$^{-1}$ of  data at 8 TeV. 
A similar sensitivity can be anticipated for 
  $Z\to \tau^\pm e^\mp$, because we
 consider   leptonic $\tau$ decays such that   $Z\to \tau^\pm \mu^\mp
\to e^\pm \mu^\mp +$ invisible. 
These limits can be compared to the  LEP1  bounds 
of order $10^{-5} \to 10^{-6}$.
Such collider searches are sensitive to a flavour-changing
effective $Z$ coupling which is energy dependent,
so are   complementary to bounds obtained from 
  $\tlll$ and $\meee$.
\vskip 10pt
\noindent
}

\end{quotation}

\vskip 20pt  

\setcounter{footnote}{0}
\renewcommand{\thefootnote}{\arabic{footnote}}


\section{Introduction}
\label{intro}

Lepton flavour is conserved in the Standard Model (SM) of particle
physics. However, the observation of neutrino oscillations
implies that there is Beyond-the-Standard-Model physics
which changes lepton flavour. 
Lepton Flavour Violation (LFV) among charged leptons is
therefore a phenomenologically motivated   place to
search for elusive Beyond-the-Standard-Model (BSM) physics.

The LHC is  an electroweak-scale hadron collider, so
can reasonably  be expected to give information on  electroweak
symmetry breaking, weakly interacting dark matter candidates
(WIMPs) and new particles or interactions in the quark
sector.  In this paper, we are interested in the prospects
of   Lepton Flavour Violating processes at the LHC,
involving SM particles. LFV in BSM scenarios, such as
supersymmetry  or heavy singlet neutrinos, has received more
attention \cite{ybook,SUSYLFV-th,SUSYLFV-ex,N@LHC}.

The LHC produces many $\tau$s in the decays of
$W$ and  $Z$ bosons, and $b$s (respectively $\sim 10^8$, 
$3\times  10^7$, and
$10^{12}$ for 10 fb$^{-1}$ at 7 TeV
\cite{t3mCMS}), so one could
hope to be sensitive to LFV $\tau$ decays
such as $\tlg$. 
LHCb obtained  a bound
$BR(\tmmm) < 7.8\times 10^{-8}$   with 1 fb$^{-1}$ of 
data \footnote{To be compared with  the Belle bound in Table~\ref{tab:L3l}.}~\cite{LHCbtmmm}.
However,
for  ATLAS and
CMS,   there  do not seem to be  enough high-$p_T$  $\tau$s, 
to allow an interesting
sensitivity~\cite{t3mCMS,Serin}. We therefore
explore LFV in the production of leptons ---
that is --- in LFV decays of the $Z$
\footnote{LFV in the decay of Higgs produced
at hadron colliders was discussed in \cite{gerald},
taking into account  the constraints from low
energy processes. An earlier study of the collider
prospects was \cite{these}. This subject has
recently been revisited \cite{oleg}.}.
The cross-section times branching ratio for
$pp \to Z \to \mu^+ \mu^-$ is of order a
nanobarn at the 7 or 8 TeV LHC,  so with
$20 $ fb$^{-1}$  of data at the end of 2012, 
the LHC  will have 
$2 \times 10^7/BR(Z\to \mu^+ {\mu}^-)  \sim 5\times 10^8  ~Z$s.
This can be compared to the $1.7 \times 10^7$ $Z$s at LEP1.
Effective LFV $Z$ couplings can also
be constrained by the possible $Z$ contribution to
decays such as  $\tmmm$ and $\meee$.  In section
\ref{sec:EFT}, we review   the complementarity
of the collider and low energy searches.

We study the  decays   $\Zme$ 
and $\Ztm$ (which results we extrapolate
to
$\Zte$). We focus on  leptonic $\tau$ decays,
so    the $\tau^\pm \ell^\mp$ final
state is $e^\pm \mu^\mp$ + two neutrinos. 
This is not without backgrounds, as discussed
in section \ref{sec:fonds}. The most dangerous
is $Z \to \tau^+ \tau^- \to e^\pm \mu^\mp + 4 $ 
neutrinos, so  section \ref{sec:EFT} also reviews some  kinematic
considerations, which suggest that this
background could be reduced to an acceptable level.
 In section \ref{sec:coupures}, we present our  selection criteria,
designed to maximise  sensitivity to the signal.  
The  bounds this
analysis  could obtain are in
the
final summary, which also  discusses future prospects. 


\section{Theoretical review}  
\label{sec:EFT}

The three $Z$ decays which violate lepton flavour conservation
are  $Z \to e^\pm\mu^\mp$, $Z \to \mu^\pm\tau^\mp$,
and $Z \to e^\pm\tau^\mp$.  We consider the
detection of $e^\pm\mu^\mp$ final states, arising  either directly,
 $Z \to e^\pm\mu^\mp$,
 or through leptonic  tau  decays: 
$Z \to \mu^\pm\tau^\mp \to \mu^\pm e^\mp\nu\bar{\nu}$, and 
$Z \to e^\pm\tau^\mp \to e^\pm \mu^\mp\nu\bar{\nu}$. 
As the last two decay channels differ only by exchange of the
electron and the muon, we only consider
 $Z \to \mu^\pm\tau^\mp \to \mu^\pm e^\mp\nu\bar{\nu}$ in this paper. 
We anticipate that the sensitivity to $Z \to e^\pm\tau^\mp $
should be equivalent, or better,  because
the  SM backgrounds should be similar to 
those of $Z \to \mu^\pm\tau^\mp \to \mu^\pm e^\mp\nu\bar{\nu}$,
and the detection prospects for  a low $p_T$
muon (from tau decay) are better than those
of a  low $p_T$ electron.

\subsection{Flavour-changing $Z$ couplings}

We suppose that  new LFV  particles 
arise  at a scale $M >m_Z$, which
allows to describe
 the  flavour changing $Z$ vertices 
with   Effective Field Theory,
via 
  operators of dimension 
six and higher\cite{Ana}\footnote{
Our discussion is patterned on
\cite{Ana}; additional references can be found
in the references and citations thereof.}.  
The discussion applies to any
flavour changing vertex, although  we focus 
on the $Z \tau^+ {\mu}^-$
interaction to avoid flavour index sums.
At dimension six, Buchmuller and Wyler\cite{BW}
give the following  SM gauge invariant operators
which can contribute (with their hermitian conjugates) to   $\Ztm$:
\bea
[H^\dagger  \sigma^I D_\a H ]
[\overline{\ell}_\mu  \sigma^I \gamma^\a \ell_\tau] ~~,~~ 
[H^\dagger D_\a H ][\overline{\ell}_\mu  \gamma^\a \ell_\tau] ~~,~~
[H^\dagger D_\a H] \overline{\mu}  \gamma^\a \tau  \label{nonop}\\
\overline{\ell}_\mu \sigma^I \gamma_\b D_\a \ell_\tau W^{I \a \b} ~~,~~
\overline{\ell}_\mu   \gamma_\b D_\a \ell_\tau B^{\a \b} ~~,~~
\overline{\mu}   \gamma_\b D_\a \tau B^{\a \b}
 \label{BWop} \\
\overline{\ell}_\mu \sigma^I H \sigma_{\b \a}  \tau W^{I \a \b} ~~,~~
\overline{\ell}_\mu H   \sigma_{\b \a}   \tau B^{\a \b} 
\label{chiralop}
\eea
where SU(2) contractions are indicated
in square brackets when there is more than one
possbility,   $\ell_\mu$ is the muon doublet,
$\tau$ the tau singlet, $B^\a$ the hypercharge
gauge boson, $\sigma^I$ are
the Pauli matrices, 
$\sigma_{\b \a} = \frac{i}{2}
[\g_\b, \g_\a]$, $B^{\a \b} = \partial^\a B^\b-\partial^\b B^\a $
and so on. We suppose that these operators are normalised such that,
{ in the presence of}
 electroweak symmetry breaking,  the combination of
operators on the first (or second, third) line of equation~(\ref{nonop} -\ref{chiralop}) 
appear in the Lagrangian as the first (or second, third) line below:
\bea
g_Z m_Z^2 [\overline{\mu} \gamma_\a(A_{L \mu \tau}P_L + A_{R \mu \tau}P_R) Z^\a \tau + h.c.]
\label{non}\\
+  2 g_Z  [\overline{\mu} \gamma_\a(C_{L \mu \tau}P_L + C_{R \mu \tau}P_R) \partial^\b \tau + h.c.] Z^{\b \a}
\label{oui}\\
+ i g_Z m_\tau [\overline{\mu} \sigma_{\a\b}(D^Z_{L \mu \tau}P_L + D^Z_{R \mu \tau}P_R)  \tau - h.c.] Z^{\a \b} 
\label{chiral}
\eea
where $g_Z = \sqrt{g^2 + g^{'2}}$, $Z^{\a \b} =
\partial^\a  Z^\b - \partial^\b  Z^\a$, and the $A,C,D$
coefficients have mass dimension -2.
 We  assume
that these interactions are generated by a loop
involving particles of mass $\sim 
M >m_Z$, so  $A,C,D \propto g_Z^2/(16 \pi^2 M^2)$.
Concretely, if $C_{L \mu \tau}, C_{R \mu \tau}$ are taken real, and the
$Z$ is on-shell, then equation~(\ref{oui}) and the SM $Z$ couplings
give 
 \beq
\frac{-g}
{2 \cos \theta_W}
\left( \!
\begin{array}{cc}
\overline{\tau} &
\overline{\mu} 
\end{array}
\!
\right)
\gamma^\a
 Z_\a  
\left[
\!
\begin{array}{cc}
g_V - g_A\gamma_5 &  
\! m_Z^2[C_{L \mu \tau} + C_{R \mu \tau}
-(C_{L \mu \tau} - C_{R \mu \tau}) \gamma_5]\\
 m_Z^2[C_{L \mu \tau} + C_{R \mu \tau}
-(C_{L \mu \tau} - C_{R \mu \tau}) \gamma_5] 
\! &  g_V - g_A\gamma_5
\end{array}
\!
\right]
\left(
\!
\begin{array}{c}
\tau\\
\mu
\end{array}
\!
\right)
\eeq
where $g_V= -1/2 + 2 s_W^2$ and  $ g_A = -1/2$. 

The effective interactions relevant to 
LFV $Z$ decays  are those 
of equation~(\ref{oui}). This is because  {
the
$A$ and $D$ coefficients are
constrained by  low energy data  (see below),
whereas }  the coefficients $C_{L},C_R $ 
are multiplied by the $Z$ four-momentum-squared,
which suppresses their contribution  in  low-energy tree\footnote{We
thank Micheal Peskin for pointing out that this suppression may
not occur in loops.} processes. For
instance,  the effective flavour-changing coupling on
the $Z$ pole would be $g_Z m_Z^2(C_{L \mu \tau} P_L + C_{R \mu \tau} P_R)$,
whereas in $\tmmm$ it would be $\lsim g_Z m_\tau^2
(C_{L \mu \tau} P_L + C_{R \mu \tau} P_R)$.  Their contribution
 to $Z$ decays is:
\bea
{BR(\Ztm)} \simeq  16 m_Z^4 (|C_{L \mu \tau}|^2  +
|C_{R \mu \tau}|^2 ) \times  {BR(\Zmm)}  \sim 1.7 \times  10^{-5} \frac{m_Z^4}{M^4}
\label{BRZ}
\eea
where  $g_V  \to 0$ and the last estimate is for
$C_{L \mu \tau} \simeq  C_{R \mu \tau} \simeq g_Z^2/(16 \pi^2M^2)$. 
Recall that the effective operator formalism supposes
$m_Z<M$, so constraints $\lsim 10^{-5}$
are interesting. 
The  current bounds on the $C$s
which can be extracted from the OPAL and DELPHI bounds on   
LFV $Z$ decays~\cite{OPALLFV,DELPHILFV} 
are  given in Table~\ref{tab:ZLFV}.
\begin{table}[htp!]
\begin{center}
\begin{tabular}{|c|c|c|}
\hline
process & exp. limit&$16 \pi^2 m_Z^2C/g_Z^2 <$ \\
\hline
$BR(Z \to e^\pm \mu ^\mp)$    (a) & $1.7 \times 10^{-6}$&0.32\\
$BR(Z \to e^\pm \tau ^\mp)$   (a) & $9.8 \times 10^{-6}$&0.76\\
$BR(Z \to \mu^\pm \tau^\mp)$  (b) & $1.2 \times 10^{-5}$&0.84\\
\hline
\end{tabular}
\caption{ Bounds on lepton flavour changing $Z$ decays 
(a) from OPAL~\cite{OPALLFV} and (b) from DELPHI \cite{DELPHILFV}. 
The third colomn assumes the decay $Z \to l^\pm_\a l^\mp_\b$ is mediated by
 $C_{L \a \b} \simeq  C_{R \a \b} \simeq
g_Z^2/16 \pi^2M^2$, and gives the resulting bound on $ m_Z^2/M^2$(which is different  for the various  $l^\pm_\a l^\mp_\b$). }
\label{tab:ZLFV}
\end{center}
\end{table}

Bounds on the $C$ couplings can also be obtained from
the contribution of a $Z$ loop to  decays such as
 $\tmg$. To { estimate} a bound, we assume that
$M \gg m_Z$, so the effective LFV vertices can
be treated as contact interactions in the loop,  and
we suppose that at the scale $M$, New Physics induces
 $C \sim g_Z^2/16 \pi^2 M^2$  and no $A$ or $D$ coefficients. 
Then, renormalisation group mixing { from the scale  
$M$  to $m_Z$ } can induce $D^\gamma$ coefficients, which we
estimate (for  the case of $\tmg$) to be
\beq
e m_\tau D^\gamma_{\mu \tau}
\sim    e m_\tau
\frac{ g_Z^2 C_{\mu \tau}^Z}{32 \pi^2} \log \frac{m_Z^2}{M^2}
\sim    e m_\tau 
\left(\frac{ g_Z^2 }{16 \pi^2}\right)^2
\frac{1 }{2M^2}
 \log \frac{m_Z^2}{M^2} ~~.
\label{CaD}
\eeq
The last
colomn of~Table~\ref{tab:L3l}
gives  bounds on the   $D^\gamma_{\a \b}$s,  which can
translated into bounds on the  $C^Z_{\a \b}$ coefficients
via equation~(\ref{CaD}). { The $\meg$ decay 
imposes  $m_Z^2 C^Z_{e \mu } \lsim  10^{-5}$, which 
corresponds to $BR(Z \to e\pm \mu^\mp) \lsim 10^{-10}$.}
 The 
 $\tlg$ decays allow
$M \sim m_Z$,  so the loop calculation
makes little sense, and  the best bounds are
the LEP limits of table~\ref{tab:ZLFV}.

We can make conflicting guesses for the ``probable''
values of $C_L$ and $C_R$.  If the new particle mass scale
$M$ is beyond the reach of the LHC, then so are 
the LFV $Z$ branching ratios  (see equation~\ref{BRZ}). Also, as discussed below,
the $A$ and $D$ coefficients are
constrained by  low energy rare decay data 
to be smaller  than  the  LEPI bounds on the $C$s.
It is unclear to the authors how to build a model
\footnote{ $A$ could be suppressed with
respect to $C$ if
the loop particles do not couple to the Higgs,
but this does not suppress  $D$  couplings.}
with $A,D \ll C$.
However, the $C$s are physically distinct,
so it is phenomenologically interesting to
constrain them. 
Furthermore,  the OPAL experiment
observed a $e^+ e^- \to e^+ \mu^-$
event at LEP2~\cite{OPAL}, 
which increases the interest
of  high centre-of-mass-energy 
 searches for LFV. 

{ To obtain  bounds from processes  at scales $\ll m_Z$, 
the $Z$ and its interactions (\ref{non}-\ref{chiral}) can be matched onto
LFV four fermion operators. }
Current rare lepton decay bounds
constrain   the  coefficients 
 $A_{L \mu \tau}, A_{R \mu \tau} $ 
  more strictly \footnote{
The operators inducing $A$ coefficients 
also induce flavour-changing four fermion
operators involving neutrinos. These are  refered to as ``Non-Standard neutrino
Interactions'', and can appear linearly in probabilities.
Despite this potential enhanced sensitivity, the
bounds   \cite{BBetal}  on  $16\pi^2 A m_Z^2/g_Z^2$ 
 are  generically $ \gsim 1$.}
 than the $Z$ decay
bounds of Table~\ref{tab:ZLFV}. The
branching ratios mediated by the $A$ coefficients are 
\bea
\frac{BR(\Ztm)}{BR(\Zmm)} \simeq  \frac{m_Z^4}{s^4_W} (|A_{L \mu \tau}|^2  +
|A_{R \mu \tau}|^2 )
~~~~,~~~
\frac{BR(\tmmm)}{BR(\tmnnb)} =
 \frac{m_Z^4}{8}
 (2 |A_{L \mu \tau}|^2  +
|A_{R \mu \tau}|^2 )
\label{BRt}
\eea
so  the rare lepton decay  bounds  of Table~\ref{tab:L3l} put $A_L$ and $A_R$ at or
beyond the sensitivity of the 7 or 8 TeV LHC.  

We neglect the $D$ coefficients of equation~(\ref{chiral}),
because their contribution to $BR(\Ztm)$ is proportional 
to $m_\tau^4$.  This is different  from
the 
photon dipole operator
$$ i e m_\tau 
[\overline{\mu} \sigma_{\a\b}(D^\gamma_{L \mu \tau}P_L + D^\gamma_{R \mu \tau}P_R) 
 \tau - h.c.] F^{\a \b} $$  where the $m_\tau^4$
factor is cancelled 
in  $BR(\tmg)$ by a similar factor  in
the denominator. 
{
The $SU(2) \times U(1)$ gauge invariant operators
of equation~(\ref{chiralop}) induce both $D^\gamma$ and
 $D^Z$, 
 so   $D^\gamma \sim D^Z$ could be expected,   in which case
radiative lepton
decays are more sensitive probes of dipole operators
than $Z$ decays. In Table~\ref{tab:L3l}, are given 
the current bounds on $D^\gamma_L = D^\gamma_R$, normalised
to $g_Z^2/16 \pi^2 m_Z^2$.}

\begin{table}[htbp!]
\begin{center}
\begin{tabular}{|c|c|c|c|}
\hline
process & exp. limit& $16 \pi^2 m_Z^2A/g_Z^2 <$&$16 \pi^2 m_Z^2D/g_Z^2 <$\\
\hline
$BR(\teee)$ &$2.7 \times 10^{-8}$ \cite{Belletmmm}&0.17& \\
$BR(\tmmm)$ &$2.1 \times 10^{-8}$ \cite{Belletmmm} &0.15&\\
$BR(\meee)$ &$ 1.0 \times 10^{-12}$ \cite{SINDRUM}&$4.3 \times 10^{-4}$&\\
$BR(\teg)$ &$3.3 \times 10^{-8}$ \cite{BelleBabar} &&$1.3 \times 10^{-3}$\\
$BR(\tmg)$ &$4.4 \times 10^{-8}$ \cite{BelleBabar} &&$1.5 \times 10^{-3}$\\
$BR(\meg)$ &$2.4 \times 10^{-12}$ \cite{MEG} &&$4.6 \times 10^{-6}$\\
\hline
\end{tabular}
\caption{ Bounds from  low energy Lepton Flavour Violating  decays,
on the $A$ and $D$ coefficients of equations~(\ref{non}) and~(\ref{chiral}).
 The third [fourth] colomn 
 assumes the decay is
mediated by $A_{L \mu \tau} \simeq A_{R \mu \tau} \simeq g_Z^2/16 \pi^2M^2$
 [$D^\gamma_{L \mu \tau} \simeq  D^\gamma_{R \mu \tau} \simeq g_Z^2/16 \pi^2M^2$], 
and gives the
bound on 
 $ m_Z^2/M^2$. If  $eD^\gamma \sim g_ZD^Z $, 
the $D^\gamma$  bounds would apply to $D^Z$. }
\label{tab:L3l}
\end{center}
\end{table}

\subsection{ Kinematics}
The LHC has more $Z$s than LEP1, but a less clean environment. 
So we here review some simple kinematics, for the
decays $\Ztl$,  which suggest that
the LHC backgrounds could be sufficiently  reduced  to
probe couplings below the LEP1 bounds.  The 
analysis for $\Zme$ is relatively straightforward, 
since the $\mu^\pm e^\mp$ should have the invariant
mass of the $Z$,
and is performed in  section \ref{subsec:emu}.

In the case of $\Ztl$,
we consider  leptonic $\tau$ decays, in particular $\tennb$,
so as to obtain opposite flavour charged leptons in the final
state:
\beq
\Ztm \to (e^\pm \nu \overline{\nu}) \mu^\mp ~~.
\label{2n}
\eeq
The various backgrounds are studied in section \ref{sec:fonds};
here we focus on the most problematic:
\beq
\Ztt \to (e^\pm \nu \overline{\nu}) (\mu^\mp \nu \overline{\nu}) ~~,
\label{4n}
\eeq
which is refered to as
``background'' for the remainder of
this section, and labelled   $Z$ + jets in Tables~\ref{tab1},~\ref{tab22}
and~\ref{tab32}. 
To estimate the number of such events,
recall that $BR(\Ztt) = 0.0337$ 
and $BR(\tennb) = 0.178$ 
\cite{PDB}.

  With some simple
approximations, one can show that
 the two-neutrino signal events
of equation~(\ref{2n})
may be kinematically distinguishable 
 from the four-neutrino
background events.  First, since the $\tau$s are
boosted by a factor of at least $m_Z/2m_\tau$, it
is not unreasonable to approximate as collinear  all
the momenta of the  $\tau$-daughters. For clarity,
suppose that the $\tau^+$  in the signal
 and background decays of equations~(\ref{2n}) and~(\ref{4n})
decays to $e^+$. Then in both decays: 
\beq
p_{\tau^+} = p_{e^+} + p_\nu + p_{\bar{\nu}} \equiv \alpha p_{e^+} ~~.
\eeq
Similarly, for  the $\tau^-$ of
the background process, one can define
 $p_{\tau^-} = 
\beta p_{{\mu}^-}$ (where we neglect all final state
masses). This gives: 
\bea
m_Z^2 -  m_\tau^2  & =&
2 \alpha   p_{e^+} \cdot p_{{\mu}^-}
~~~~~~~~~~~~~{\rm signal } \label{a1}  \\
m_Z^2 - 2 m_\tau^2  & =&
2 \alpha\beta   p_{e^+} \cdot p_{{\mu}^-}
~~~~~~~~~~~{\rm background } 
\eea
As expected, the invariant mass $2 p_{e^+} \cdot p_{{\mu}^-}$
 of the charged final
state leptons should be lower for the background events,
and the $p_T$ of the signal muon will have
a harder distribution than that of the $\tau$ daughters (see
Fig.~\ref{fig1}).
 
The second approximation is to  neglect
the transverse momentum $p_{T,Z}$ of the $Z$,
so that the $p_T$ of the $Z$-daughters balance. This 
 implies that 
\bea
\alpha |p_{T,e^+}| = |p_{T,\mu^-}| ~~~~~~~~~~~~~~~~~~~~~{\rm signal}
\label{a2}   \\
\alpha |p_{T,e^+}| = \beta |p_{T,\mu^-}| ~~~~~~~~~~~~{\rm background } 
\eea
To distinguish signal and background, we 
extract $\alpha_1$ and $\alpha_2$  from 
our ``data'' using 
respectively  equations~(\ref{a1})
and~(\ref{a2}), and use the difference
\beq
\Delta \alpha = \alpha_2 - \alpha_1
\label{Da}
\eeq
as a discriminating variable between background
and  signal (which peaks  at zero).

There is  a   small fraction of
background  events, with $\beta \to 1$, 
 which constitute  an irreducible background
to our signal.
These correspond to events
 where  the $\mu$  carries (almost) all the momentum
of its parent-$\tau$.  They   will occur when, in the
reference frame of the decaying $\tau$, the $\mu$ 
carries close to the maximum allowed energy 
$(m_\tau^2 + m_\mu^2)/2m_\tau$, and when it is emitted
in the direction of the boost of the $\tau$.
If the charged lepton energies are measured with
a  fractional uncertainty $\epsilon$, then 
we estimate that such irreducible events correspond
to $\sim \epsilon^2$ of the background. For
$\epsilon \sim$ few \%, this suggests that
the background  of equation~(\ref{4n}) can be 
sufficiently reduced.



\section{Background and signal simulation}
\label{sec:fonds}

The signal investigated in this paper consists of events with exactly one electron and one muon.
At the LHC, the dominant SM process leading to $e^{\pm}\mu^{\mp}$ dilepton final states 
is $Z/\gamma^* \to \tau^+\tau^-$ where both tau lepton decays leptonically, one 
to a muon and neutrinos and the
other  to an electron and neutrinos. Two processes involving
top quark(s) also lead to $e^{\pm}\mu^{\mp}$ final
states: the top quark pair production channel 
$pp \to t\bar{t} \to b\ell^{+}\nu \bar{b}\ell'^-\bar{\nu}$ and the single-top production channel 
$pp \to Wt \to \ell^{\pm}\nu b\ell'^\mp\nu$
with $\ell,\ell'=e,~\mu,~\tau$~\footnote{The notation $pp \to Wt$ includes both $pp \to tW^-$ and $pp \to \bar{t}W^+$.}. 
Finally, gauge boson pair production
can lead to
$e^{\pm}\mu^{\mp}$ states  via: $pp \to W^+W^- \to \ell^+\nu \ell'^-\bar{\nu}$ and 
$pp \to W Z/\gamma^* \to \ell'\nu \ell^+\ell^-$ with 
$\ell,\ell'=e,~\mu,~\tau$, $pp \to W Z/\gamma^* \to q\bar{q}'\tau^+\tau^-$, and $pp \to Z/\gamma^*Z/\gamma^* \to f\bar{f}f'\bar{f'}$
with $f=q,\nu,e,~\mu,~\tau$.

The {\sc fewz}~\cite{Gavin:2010az} program was used to compute the inclusive $pp \to Z/\gamma^*$ cross section at 
next-to-next-to-leading order (NNLO) at $\sqrt{s}=8~TeV$. The next-to-next-to-leading logarithm (NNLL) cross sections were taken from 
Ref.~\cite{Cacciari:2011hy} for $pp \to t\bar{t}$, and from Ref.~\cite{Kidonakis:2010ux} for $pp \to tW$, with a 
top quark mass of 173.3~GeV. Finally, the diboson ($pp \to WW,~WZ,~ZZ$) cross sections were computed using {\sc mcfm}~\cite{Campbell:2011bn} 
at next-to-leading order (NLO).These results are used to predict the
total number of background events, which we then simulate at leading order.

The simulation of those SM backgrounds
and of the signal processes  $pp \to Z \to \mu^{\pm}\tau^{\mp} \to \mu^{\pm} e^{\mp}\nu\bar{\nu}$ 
and $pp \to Z \to e^{\pm}\mu^{\mp}$  was performed  using {\sc pythia} version~8.162~\cite{Sjostrand:2007gs}, 
except the simulation of the $pp \to Wt \to \ell^{\pm}\nu b\ell'^\mp\nu$ process
which also  used {\sc madgraph5}
version~1.4.6~\cite{Alwall:2011uj}. In all Monte Carlo simulations, the {\sc cteq6l1}~\cite{cteq6} parton density 
functions were used; initial and final state radiations and multi-parton interactions were also included.
However, pile-up effects due to additional interactions were not taken into account in those simulations.

All Monte Carlo events were passed through the {\sc delphes}~\cite{Ovyn:2009tx} program
which provides a fast simulation of LHC detector response. 
The default
 configuration corresponding to the CMS detector~\cite{Bayatian:2006zz} at the LHC was used.
In the following, the pseudorapidity is defined as $\eta=-\ln \left[ \tan \left( \theta/2 \right) \right]$
with $\theta$ the polar angle with respect to the beam direction; and $\phi$ is the azimuthal angle.
Isolated electrons and muons were identified by requiring no track with $p_T>2$~GeV
in a cone of radius $R=0.5$ around those lepton candidates, with $\Delta R = \sqrt{\Delta\eta^2+\Delta\phi^2}$. 
Standard Model backgrounds with non-isolated leptons
were therefore not taken into account. Jets were reconstructed with the {\sc fastjet}~\cite{Cacciari:2011ma}
program using the anti-kt algorithm with a distance parameter of $0.7$.

Table~\ref{tab1} contains the details of the Monte Carlo event production performed for the studies reported in 
this paper. As a total integrated luminosity of $20~fb^{-1}$ at $\sqrt{s}=8~TeV$ is assumed in the following analyses,
the equivalent luminosities of these simulations are always higher than $200~fb^{-1}$ for each process to reduce statistical
uncertainties.

\begin{table}[htb]
\begin{center}
\begin{tabular}{|c|l|r|r|}
\hline
\multicolumn{2}{|c|}{Processes}                          & $\sigma \times BR$ (pb) & Number of \\
\multicolumn{2}{|c|}{\ }                                 &                         & simulated events              \\
\hline\hline
$Z+$jets   & $Z/\gamma^* \to \tau^{\pm}\tau^{\mp} \to \ell^{\pm}\nu\bar{\nu} \ell'^{\mp}\nu\bar{\nu}$ with $\ell,\ell'=e,\mu$ ($M_{Z/\gamma^*}>20$~GeV)        & 237.    & 47,500,000 \\
\hline
$t\bar{t}$ & $t\bar{t} \to b\ell^{+}\nu \bar{b}\ell'^-\bar{\nu}$ with $\ell,\ell'=e,\mu,\tau$                         &	 23.6   &  4,720,000 \\
\hline
single-top & $Wt \to \ell^{\pm}\nu b\ell'^\mp\nu$ with $\ell,\ell'=e,\mu,\tau$                                        &   2.35  &    400,000 \\
\hline
diboson  & $W^+W^- \to \ell^+\nu \ell'^-\bar{\nu}$ with $\ell,\ell'=e,\mu,\tau$                                       &	  6.00  &  1,200,000 \\
 & $WZ/\gamma^* \to q\bar{q}'\tau^{\pm}\tau^{\mp}$					                              &	  0.735 &    140,000 \\
 & $WZ/\gamma^* \to \ell'^{\pm}\nu \ell^+\ell^-$ with $\ell,\ell'=e,\mu,\tau$	                                      &	  1.06  &    220,000 \\
 & $Z/\gamma^*Z/\gamma^* \to f\bar{f}f'\bar{f}'$ with $f=q,\nu,e,~\mu,~\tau$					      &	  8.26  &  1,660,000 \\
\hline
signals & $Z \to \tau^{\pm} \mu^{\mp} \to e^{\pm}\nu\bar{\nu} \mu^{\mp}$                                              &          & 100,000 \\ 
        & $Z \to \mu^{\pm} e^{\mp}$							                              &          & 100,000 \\
\hline
\end{tabular}
\caption{\label{tab1}
Standard model backgrounds in the search for $Z$ decay violating lepton number at the LHC;
the second column contains $\sigma \times BR$ (in pb) for p-p collisions at $\sqrt{s}=8~TeV$,
and the third column shows the number of simulated events in this analysis which is in all cases 
greater than 10 times $\mathcal{L}\sigma BR$ with  $\mathcal{L}=20~fb^{-1}$.
}
\end{center}
\end{table}


\section{Sensitivity at the 8~TeV LHC}
\label{sec:coupures}

\begin{figure}[htbp]
\begin{center}
\includegraphics[width=8cm]{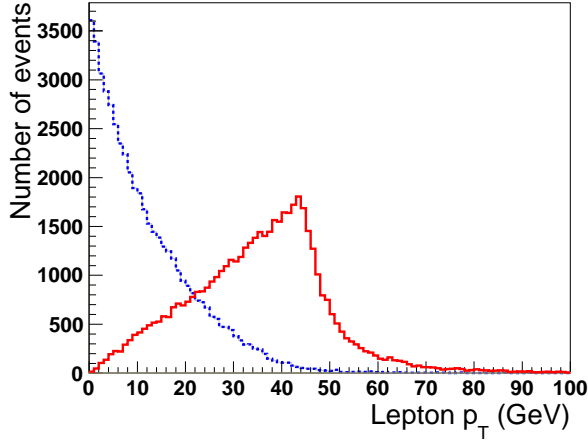}
\caption{
\label{fig1}
Lepton (electron or muon) $p_T$ distribution at generator level at the 8~TeV LHC: for leptons 
from $pp \to Z/\gamma^* \to \ell^+\ell^-$ (plain line) and for leptons from 
$pp \to Z\gamma^* \to \tau^+\tau^- \to \ell^+\nu\bar{\nu}~\ell^-\bar{\nu}\nu$ (dotted line), with $\ell=e$~or~$\mu$.
}
\end{center}
\end{figure}

For the signal investigated in this study, at least one lepton comes directly from the 
$Z$ boson decay. As shown in Fig.~\ref{fig1}, the distribution of this lepton $p_T$
has a clear peak around $\sim~40$~GeV, as expected for this jacobian distribution.
At the LHC, a large fraction of SM backgrounds leading to $e^\pm\mu^\mp$ final states
comes $W$ and $Z$ bosons decaying to tau lepton(s) which decay leptonically 
($\tau \to \ell \bar{\nu} \nu$). In that case, the $p_T$ of the final lepton is strongly 
reduced due to the energy carried away by neutrinos (see Fig.~\ref{fig1}).
Requiring at least one high $p_T$ lepton,
can significantly increase the signal to background ratio.
This selection criterium also guarantees that events are
accepted by single lepton trigger conditions of LHC experiments. In the following, 
100\% trigger efficiency is therefore assumed.

\subsection{The $\bm{pp \to Z \to \mu^{\pm}\tau^{\mp}}$ channel}
\label{subsec:taumu}

In the $pp \to Z \to \mu^{\pm}\tau^{\mp} \to \mu^\pm e^{\mp}\bar{\nu}\nu$ channel, the muon comes 
directly from the $Z$ boson decay. The muon $p_T$ is required to be above 30~GeV and in the detector
pseudorapidity acceptance $|\eta|<2.1$. As the electron comes from the tau decay, the selection criterium
on the electron $p_T$ has to be reduced to keep high signal efficiency. The electron $p_T$ is therefore
required to be above 10~GeV and with $|\eta|<2.5$. To reduce the diboson background coming from events with 
more than 3 leptons, events are required to contain exactly 2 leptons. At this stage, a reducible diboson 
background comes from events where lepton(s) failed identification criteria leading to same-sign 
$e^\pm\mu^\pm$ final states. This background is therefore rejected by requiring two opposite-sign leptons.
Double and single-top quark production are then strongly reduced by rejecting events which contain
at least one jet with $p_T>30$~GeV and $|\eta|<2.5$. To reduce the remaining diboson backgrounds, 
angular criteria in the transverse plane are used. For the signal, electrons and muons are 
almost back-to-back in the transverse plane, while the distribution of the $\Delta\phi$ angle between 
the muon and the electron is more flat for the diboson backgrounds. This $\Delta\phi(e,\mu)$ angle 
is therefore required to be higher than 2.7 radians. Another difference between backgrounds and signal
is the $\Delta\phi$ angle between the electron and the direction of the missing transverse energy ($\met$),
which is peaked at 0 for signal since electrons and missing transverse energy
both come from a higly boosted tau lepton. This $\Delta\phi(e,\met)$ angle is therefore required to be 
below 0.7 radians. Finally, the $e\mu$ invariant mass is required to be in the range $[38-92]$~GeV, i.e.
in the region around the broad peak expected for the signal. The numbers of events expected from SM 
backgrounds at each step of the event selection are reported in Table~\ref{tab21}, together with the signal
efficiency.

\begin{table}[htb]
\begin{center}
\begin{tabular}{ | l | r | r | }
  \hline
  Selection criteria                                        & $N_{backgrd.} $  & Signal efficiency (\%) \\
  \hline\hline
  $\geq$~1 muon with $p_T>30$~GeV and $|\eta|<2.1$ and      & 48,181 $\pm$ 68 &  9.4 $\pm$ 0.1 \\
  $\geq$~1 electron with $p_T>10$~GeV and $|\eta|<2.5$      &                &                  \\
  \hline
  exactly 2 leptons                                         & 43,496 $\pm$ 65 &  9.4 $\pm$ 0.1 \\
  2 leptons with opposite charge                            & 42,652 $\pm$ 64 &  9.4 $\pm$ 0.1 \\
  jet veto: no jet with $p_T>30$~GeV and $|\eta|<2.5$       & 11,358 $\pm$ 34 &  7.8 $\pm$ 0.1 \\
  $\Delta\phi(e,\mu) > 2.7$                                 &  6,850 $\pm$ 26 &  6.9 $\pm$ 0.1 \\
  $\Delta\phi(e,\met) < 0.7$                                &  3,763 $\pm$ 19 &  6.2 $\pm$ 0.1 \\
  $38 \, GeV < M_{e\mu} < 92 \, GeV$			    &  3,201 $\pm$ 18 &  6.1 $\pm$ 0.1 \\
  \hline
\end{tabular}
\caption{\label{tab21}
Selection criteria for the $Z\to \tau^{\pm}\mu^{\mp}$ search at the $\sqrt{s}=8$~TeV LHC with $\mathcal{L}=20~fb^{-1}$
with the total number of events expected from SM backgrounds and the signal efficiency (\%); uncertainties are statistical
only.
}
\end{center}
\end{table}

The remaining SM backgrounds is 3201 events. As expected, 95\% of this background comes from 
the $Z/\gamma^* \to \tau^{\pm}\tau^{\mp} \to \mu^{\pm} e^{\mp}\nu\bar{\nu}$ background 
(see Tab.~\ref{tab22}). The signal efficiency is 6.1~\%. Normalized to the $pp \to Z$ cross section
multiplied by $BR(Z \to \tau^{\pm}\mu^{\pm})=1.2 \times 10^{-5}$, which corresponds to the current 
95\% C.L. limit from LEP, the number of events expected for the signal is 489. 

\begin{table}[htb]
\begin{center}
\begin{tabular}{ | l | r | }
  \hline
  Processes  & Number of events \\
  \hline
  \hline
  $Z+$jets   & 3,037 $\pm$ 17 \\
  $t\bar{t}$ &     9 $\pm$  1 \\
  single-top &     7 $\pm$  1 \\
  diboson    &   147 $\pm$  4 \\
  \hline\hline
  Total SM backgrounds     & 3,201 $\pm$ 18 \\
  \hline
  \hline
  $Z\to \tau^{\pm}\mu^{\mp}$ signal & 489 $\pm$ 6 \\
  \hline
\end{tabular}
\caption{\label{tab22}
After selection criteria of Table~\ref{tab21} for the $Z\to \tau^{\pm}\mu^{\mp}$ search at the $\sqrt{s}=8$~TeV LHC with $\mathcal{L}=20~fb^{-1}$,
number of events expected for each SM background and for the signal normalized to $\sigma (pp \to Z) \times BR(Z \to \tau^{\pm}\mu^{\pm})$
with $BR(Z \to \tau^{\pm}\mu^{\pm})=1.2 \times 10^{-5}$. These numbers correspond to Fig.~\ref{fig2}.
}
\end{center}
\end{table}

\subsection{The $\bm{pp \to Z \to e^{\pm}\mu^{\mp}}$ channel}
\label{subsec:emu}

In the $pp \to Z \to e^{\pm}\mu^{\mp}$ channel, selection criteria are very close to those reported in 
previous section. The fact that this signal contains two high $p_T$ leptons with a narrow invariant mass
allows to reach a much better signal to background ratio. The muon and the electron $p_T$ are required 
to be above 30~GeV and their pseudorapidity $|\eta|<2.1$ and $<2.5$, respectively. Events are required
to contain exactly two leptons, with opposite charge. The jet veto rejecting events with at least one 
jet with $p_T>30$~GeV and $|\eta|<2.5$ is then applied. Finally, the $\Delta\phi(e,\mu)$ angle is 
required to be higher than 2.7~radians. The numbers of events expected from SM backgrounds at each 
step of the event selection are reported in Table~\ref{tab31}, together with the signal
efficiency.

\begin{table}[htb]
\begin{center}
\begin{tabular}{ | l | r | r | }
  \hline
  Selection criteria                                        & $N_{backgrd.} $  & Signal efficiency (\%) \\
  \hline\hline
  $\geq$~1 muon with $p_T>30$~GeV and $|\eta|<2.1$ and      & 36,489 $\pm$ 59 & 18.5 $\pm$ 0.1 \\
  $\geq$~1 electron with $p_T>30$~GeV and $|\eta|<2.5$      &                &                  \\
  \hline
  exactly 2 leptons                                         & 34,787 $\pm$ 58 & 18.5 $\pm$ 0.1 \\
  2 leptons with opposite charge                            & 34,038 $\pm$ 58 & 18.5 $\pm$ 0.1 \\
  jet veto: no jet with $p_T>30$~GeV and $|\eta|<2.5$       &  8,111 $\pm$ 28 & 15.9 $\pm$ 0.1 \\
  $\Delta\phi(e,\mu) > 2.7$                                 &  4,771 $\pm$ 22 & 14.0 $\pm$ 0.1 \\
  \hline
\end{tabular}
\caption{\label{tab31}
Selection criteria for the $Z\to e^{\pm}\mu^{\mp}$ search at the $\sqrt{s}=8$~TeV LHC with $\mathcal{L}=20~fb^{-1}$
with the total number of events expected from SM backgrounds and the signal efficiency (\%); uncertainties are statistical
only.
}
\end{center}
\end{table}

The remaining SM backgrounds is 4771 events. At this stage the dominant background comes from
$Z/\gamma^* \to \tau^{\pm}\tau^{\mp} \to \mu^{\pm} e^{\mp}\nu\bar{\nu}$ events
(see Tab.~\ref{tab32}). And the signal efficiency is 14.0~\%. Normalized to the $pp \to Z$ cross section
multiplied by $BR(Z \to e^{\pm}\mu^{\pm})=1.7 \times 10^{-6}$, which corresponds to the current 
95\% C.L. limit from LEP, the number of events expected for the signal is 158. 
In the mass region $[87-95]$ around the $Z$ peak, this signal would give 121 events, whereas
the total SM background represents 222 events,
half of it coming $Z/\gamma^* \to \tau^{\pm}\tau^{\mp} \to \mu^{\pm} e^{\mp}\nu\bar{\nu}$ events
and the other half from diboson events.

\begin{table}[htb]
\begin{center}
\begin{tabular}{ | l | r | }
  \hline
  Processes  & Number of events \\
  \hline
  \hline
  $Z+$jets   & 2,756 $\pm$ 17 \\
  $t\bar{t}$ &   199 $\pm$  4 \\
  single-top &   132 $\pm$  4 \\
  diboson    & 1,683 $\pm$ 13 \\
  \hline\hline
  Total SM backgrounds     & 4,771 $\pm$ 22 \\
  \hline
  \hline
  $Z\to e^{\pm}\mu^{\mp}$ signal & 158 $\pm$ 1 \\
  \hline
\end{tabular}
\caption{\label{tab32}
After selection criteria of the $Z\to e^{\pm}\mu^{\mp}$ search at the $\sqrt{s}=8$~TeV LHC with $\mathcal{L}=20~fb^{-1}$,
number of events expected for each SM background and for the signal normalized to $\sigma (pp \to Z) \times BR(Z \to e^{\pm}\mu^{\pm})$
with $BR(Z \to e^{\pm}\mu^{\pm})=1.7 \times 10^{-6}$. These numbers correspond to Fig.~\ref{fig2}.
}
\end{center}
\end{table}

\subsection{Limits}
\label{subsec:limits}

\begin{figure}[htbp]
\begin{center}
\begin{tabular}{cc}
\includegraphics[width=8cm]{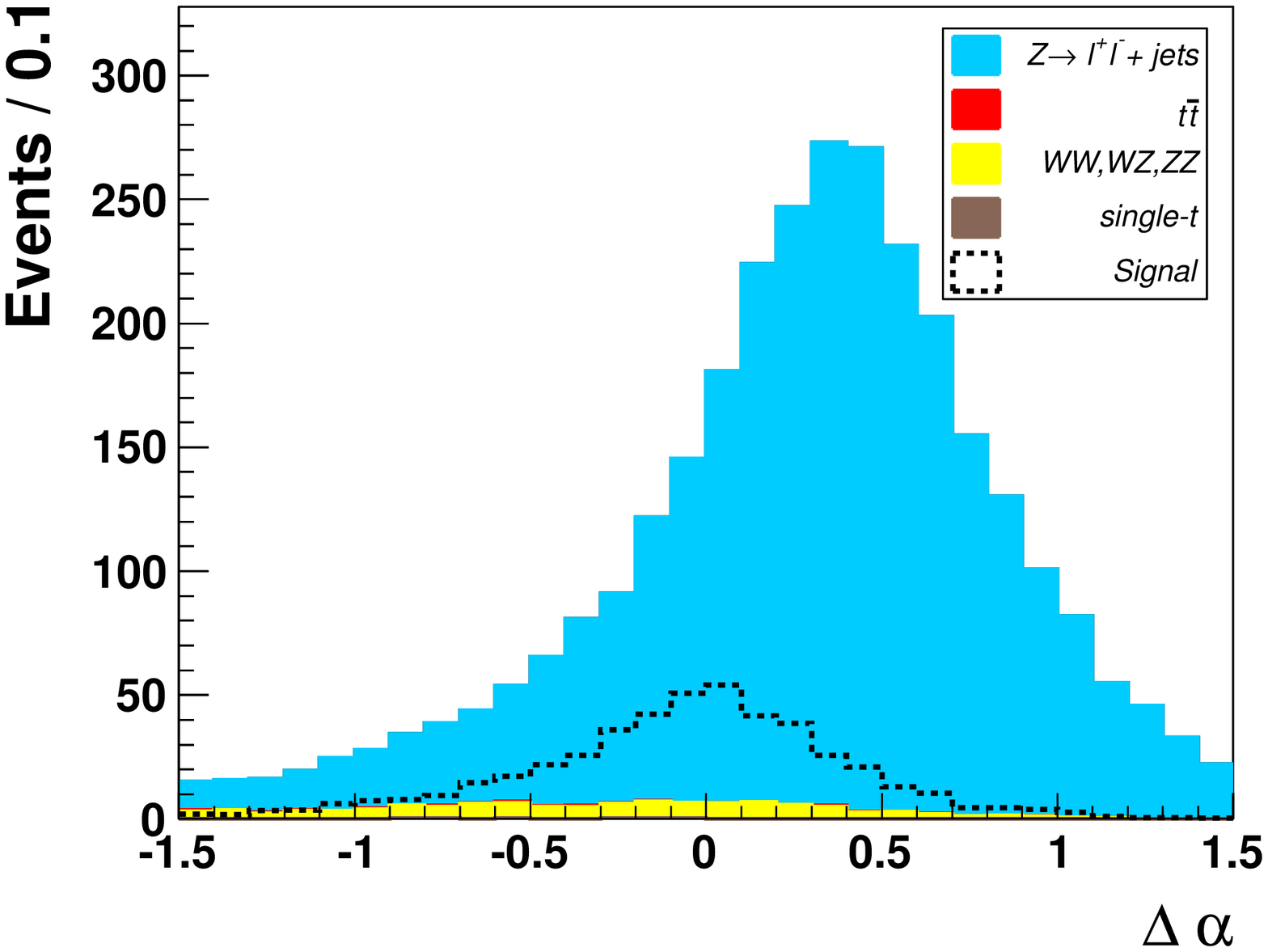} &
\includegraphics[width=8cm]{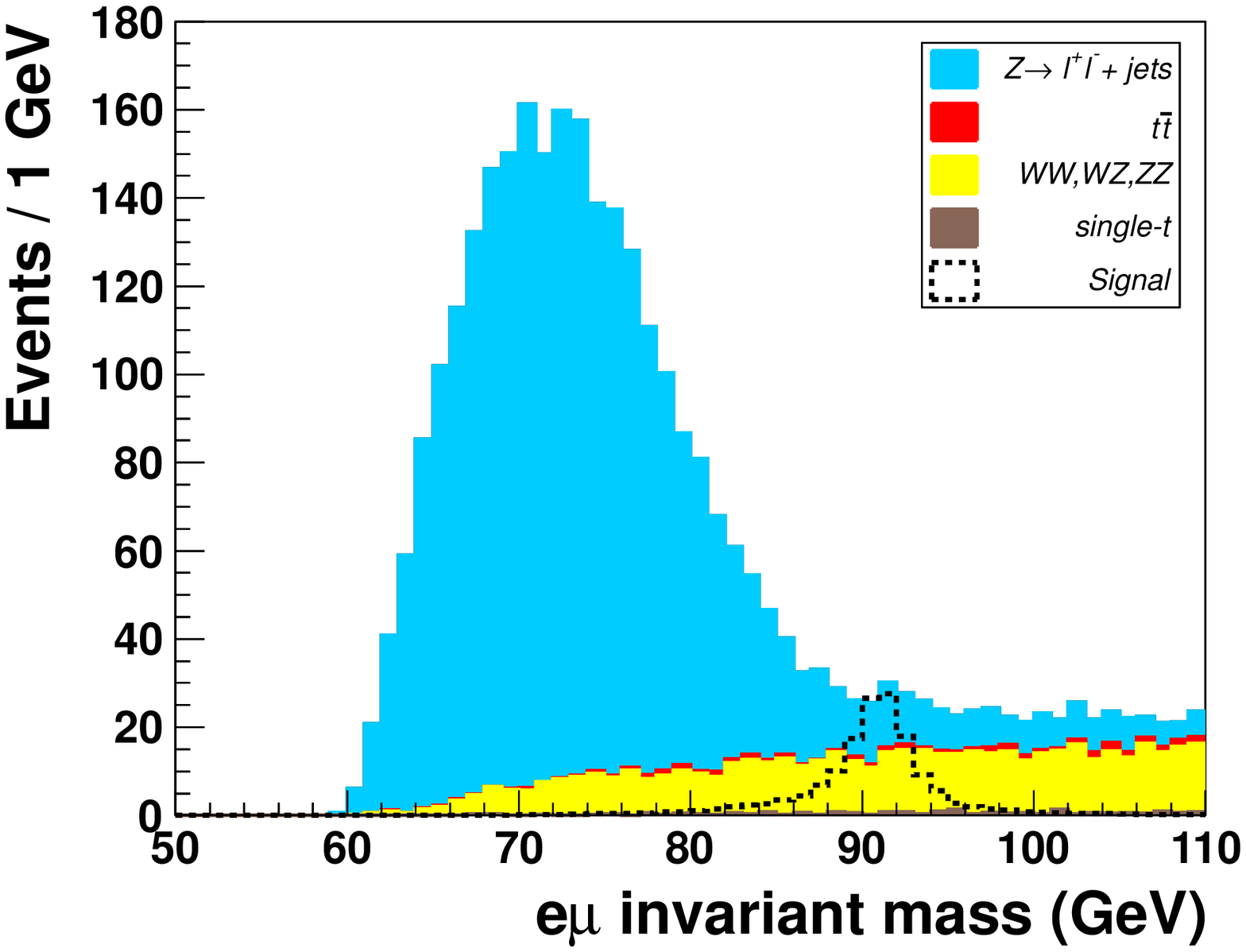} \\
\end{tabular}
\caption{
\label{fig2}
Search for flavour violating Z decay at the $\sqrt{s}=8$~TeV LHC with $\mathcal{L}=20~fb^{-1}$:
in the $Z \to \mu^{\pm}\tau^{\mp} \to \mu^{\pm} e^\mp \nu\bar{\nu}$, distribution
of $\Delta\alpha$ after the final step of the event selection (left), and
in the $Z\to e^{\pm}\mu^{\pm}$ channel, distribution of the $e^{\pm}\mu^{\pm}$ invariant mass 
after the final step of the event selection (right).
The  dashed signal  corresponding to the LEP1 limits is superposed on the summed backgrounds.
}
\end{center}
\end{figure}

From the number of background events and the signal efficiencies reported in the previous section, 
expected 95\% C.L. limits on the number of signal events can be computed and translated into limits
on lepton flavour violating $Z$ branching ratios. The modified frequentist 
$CL_s$~\cite{cls:junkread} method was used to compute those expected limits.
The shapes of the $\Delta\alpha$
(defined in equation~\ref{Da})
 and $e^\pm\mu^\mp$ invariant mass distributions were used as final 
discriminant variables in the limit computation for the $Z\to \tau^\pm\mu^\mp$ and 
$Z\to e^\pm\mu^\mp$ channel, respectively. 
Those distributions are shown in Fig.~\ref{fig2}.
A 3\% systematic uncertainty was set on the SM background contributions and on the signal
efficiencies. This uncertainty accounts for the experimental uncertainty on lepton 
identification~\cite{lhcwz}, assuming that the uncertainty on luminosity measurement
can be cancelled in our analysis using normalization to $pp \to Z \to \mu^+\mu^-$ and 
$pp \to Z \to e^+e^-$ events.
This 3\% systematic uncertainty was treated as fully correlated between signal and SM backgrounds.
This procedure allows to set 95\% C.L. expected limits of 
$3.5 \times 10^{-6}$ and $4.1 \times 10^{-7}$ on the $Z$ boson branching ratios 
$BR(Z \to \tau^\pm\mu^\mp)$ and $BR(Z \to e^\pm\mu^\mp)$, respectively. Those expected limits are a factor
$\sim$4 better than the current experimental limits from LEP.

\section{Summary}
\label{summary}

The high number of $Z$ bosons produced at the LHC allows to study rare $Z$ boson decay predicted
by BSM theories. Using $20~fb^{-1}$ of $pp$ collisions at 8~TeV, 
we have shown that 95\% C.L. 
limits on $Z$ decay branching ratios to $\tau^{\pm}\mu^{\mp}$ and 
$e^{\pm}\mu^{\mp}$ of 
$3.5 \times 10^{-6}$ and $4.1 \times 10^{-7}$, respectively, could 
be obtained. This represents
a factor 4 improvements with respect to current best limits from 
LEP. 
The expected limit on 
$BR(Z \to \tau^{\pm}e^{\mp})$ should be equivalent to the one on 
$BR(Z \to \tau^{\pm}\mu^{\mp})$.
As discussed in section~\ref{sec:EFT}, such collider 
searches are generically  complementary to
rare lepton decay searches, because they are sensitive 
to an effective vertex $\propto s$, { which gives
small  contributions to  tree level low energy processes.}
As shown in Eq.~\ref{BRZ}, reaching sensitivity to lower 
$BR$s would be interesting to test
BSM models.
{ However, 
an  effective $Z$ vertex   $\propto s$ 
can also contribute  in loops above $m_Z$;
the  limits on $\meg$   are sufficiently  restrictive
to constrain
this effect, implying that 
 $BR(\Zme) \lsim 10^{-10}$.}

In the analyses presented in this paper, basic and simple selection criteria were used to 
distinguish signals from SM backgrounds. More sophisticated analyses might further reduce 
backgrounds and increase sensitivities. In the limit computation, a 3\% total systematic 
uncertainy on signal and background has been assigned, assuming some systematics, as the one
on the integrated luminosity, could be cancelled in this measurement. 

It is interesting to speculate about the prospects of LFV $Z$
decays at the post-2014 LHC. Our analysis
is not immediately  applicable, as the relative importance of the 
backgrounds will vary with the increased centre-of-mass energy.  
However, for the analysis described here,
the expected limits  reduce  by  factors of 2 or 3 
with a  luminosity  of  $200~fb^{-1}$,
then further improvements are limited by the systematic uncertainties. 
Selection criteria  which more effectively suppress  
the backgrounds, and reduced systematic uncertainties, would be 
required to profit from the full integrated luminosity of the future LHC.


\section*{Acknowledgements}
We thank Gerald Grenier  and Michael Peskin for useful suggestions.


\end{document}